\documentclass[prl,reprint,showpacs,showkeys,showpacs,times,superscriptaddress]{revtex4-1}
\usepackage{amsmath,amssymb,amsthm,times,graphics,graphicx,bm,subfigure}

\begin{document}

\title{Experimental ancilla-assisted phase-estimation in a noisy channel}
\author{Marco Sbroscia}
\affiliation{Dipartimento di Scienze, Universit\`a degli Studi Roma Tre, Via della Vasca Navale 84, 00146, Rome, Italy}
\author{Ilaria Gianani}
\affiliation{Dipartimento di Scienze, Universit\`a degli Studi Roma Tre, Via della Vasca Navale 84, 00146, Rome, Italy}
\author{Luca Mancino}
\affiliation{Dipartimento di Scienze, Universit\`a degli Studi Roma Tre, Via della Vasca Navale 84, 00146, Rome, Italy}
\author{Emanuele Roccia}
\affiliation{Dipartimento di Scienze, Universit\`a degli Studi Roma Tre, Via della Vasca Navale 84, 00146, Rome, Italy}
\author{Zixin Huang}
\affiliation{School of Physics, University of Sydney, NSW 2006, Australia}
\author{Lorenzo Maccone}
\affiliation{Dipartimento di Fisica INFN, Sezione di Pavia, Universit\`a di Pavia, Via Bassi 6 I-27100 Pavia, Italy}
\author{Chiara Macchiavello}
\affiliation{Dipartimento di Fisica INFN, Sezione di Pavia, Universit\`a di Pavia, Via Bassi 6 I-27100 Pavia, Italy}
\author{Marco Barbieri}
\affiliation{Dipartimento di Scienze, Universit\`a degli Studi Roma Tre, Via della Vasca Navale 84, 00146, Rome, Italy}

\begin{abstract}
The metrological ability of carefully designed probes can be spoilt by the presence of noisy processes occurring during their evolution. The noise is responsible for altering the evolution of the probes in such a way that bear little or no information on the parameter of interest, hence spoiling the signal-to-noise ratio of any possible measurement. Here we show an experiment in which the introduction of an ancilla improves the estimation of an optical phase in the high-noise regime. The advantage is realised by the coherent coupling of the probe and the ancilla at the initialisation and the measurement state, which generate and then select a subset of overall configurations less affected by the noise process.
\end{abstract}
\maketitle


Optimal design of measurements for parameter estimation exploits the selection of the more informative state among the whole possible configuration space of the probe. Such a state is the one most affected by the perturbation imposed by the interaction with the sample. Strategies based on the use of quantum resources take benefit from this to deliver enhanced estimation \cite{giovannetti2004, giovannetti2006}.

In the absence of noise, the evolution characterised by the parameter establishes the number of resources to be considered: in the common example of phase estimation in the context of quantum optics implementation, we need two optical modes, one acting as a reference, and the other probing the phase. In quantum mechanical terms this means setting the dimensionality of the optimal states, but similar considerations can be drawn in the classical case. The presence of noise effectively enlarges the set of possible configurations, for instance by introducing loss modes \cite{holland1993,dorner2009,kacp2010,knysh2011,escher2011,genoni2011,genoni2012}: the effect is generally detrimental to the precision of the estimation, and appropriate strategies need to be put in place.

At the same time, it is well known that expanding the configuration space provides more flexibility in measurement design: the first demonstration concerned the implementation of unambiguous discrimination of two non-orthogonal quntum states~\cite{huttner1996}, and it has since been extended to the realisation of generalised measurements~\cite{pryde2005,mosley2006,ling2008,baek2008,saunders2012}. Furthermore, the approach of introducing ancillary systems can provide advantages in the characterisation of quantum processes~\cite{altepeter2003,chiuri2011}, and in the compact design of quantum circuits~\cite{layon2009,monz2009,zhou2011,hor2014}.

\begin{figure}[h!]
\includegraphics[width=0.4\textwidth]{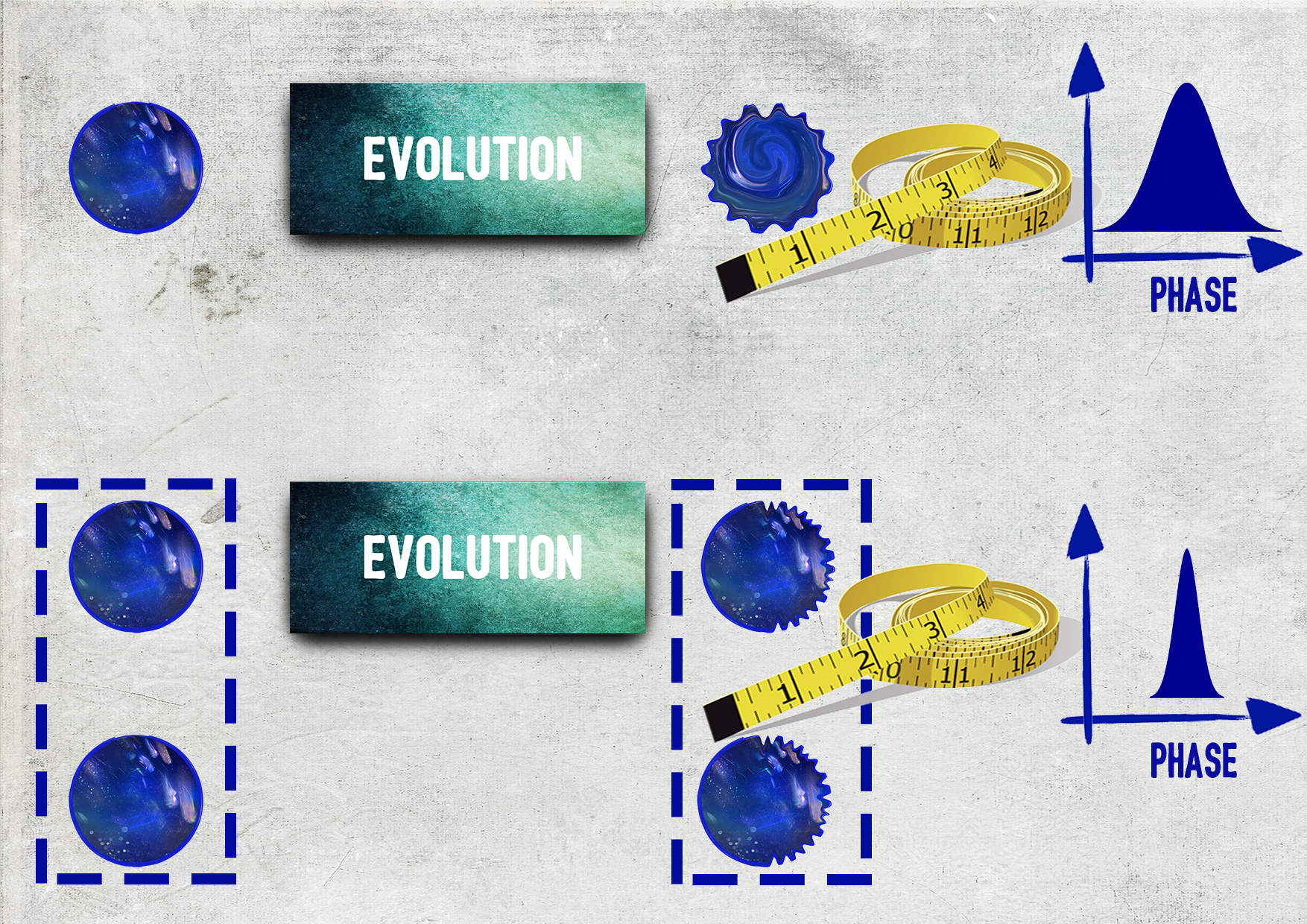}
\caption{Concept of the comparison between measurement strategies. The standard measurement addresses a single probe state which has undergone a noisy evolution, resulting in a phase uncertainty (upper panel).  Introducing a correlated ancilla, and implementing a collective measurement after the evolution gives an improved estimation (lower panel).
}
\label{fig1}
\end{figure}

Even though the use of ancillae was motivated by the realisation of generalised measuments, only recently it has been considered under the metrology perspective: we now know that introducing an ancilla is an helpful tool for parameter estimation in the presence of noise~\cite{demko2014}, see Fig. \ref{fig1}. Here we show experimentally how an ancilla-assisted strategy pushes the precision limit in phase estimation, and discuss the nature of origin of this advantage. We used single photons to probe a phase in an environment imposing amplitude damping. We could verify that coupling the polarisation of the photons with their path at both probe initialisation and readout stages resulted in improved precision in the high-noise regime. From the quantum metrology perspective, our findings support and extend previous results~\cite{girolami2014, roccia2017} that demonstrated how collective measurements offers better performance at handling noise. For classical optics, our results open a new perspective in metrological methods inspired by an analysis of the apparatus at the quantum level~\cite{dambrosio2013, altorio2015, altorio2016,thiel2017}.

Our experimental scheme implements a qubit phase estimation in the presence of amplitude damping (AD) noise by measuring a birefringent optical phase in the presence of controlled noise . We used single photons generated by Spontaneous Parametric DownConversion (SPDC) in a Type-I non-linear crystal (BBO) with a 3 mm, length pumped with an 80 mW continuous wave 405 nm-wavelenght laser. One of the two correlated photons produced in the SPDC has been sent to the interferometer, the other has been used as a trigger.

The standard procedure consists in preparing these photons in the diagonal $\vert D\rangle=1/\sqrt{2}\left(\vert H\rangle+\vert V\rangle\right)$, where H and V are the horizontal and vertical polarisation state of the photon respectively, sending them through the noisy channel and then performing the optimal measurement, identified by optimising the highest Quantum Fisher Information (QFI). Clearly, the precision will be affected by the fact that a number of probe photons will be transferred to a noisy mode that has no coherence with the original modes. This is captured by the QFI which, in the presence of a damping rate $\eta$, decreases from the noiseless value 1 to $F_s=1-\eta$~\cite{zixin2016}. Thus the damping rate $\eta$ plays the role of noise level. In our experiment we considered a different approach: the initial two-level system was coupled to an ancillary one that in our implementation corresponded to a different degree of freedom of the same photon. It is necessary that such coupling is realised in a coherent fashion; we will comment later on the exact nature of this operation. 

\begin{figure}[h!]
\includegraphics[width=0.5\textwidth]{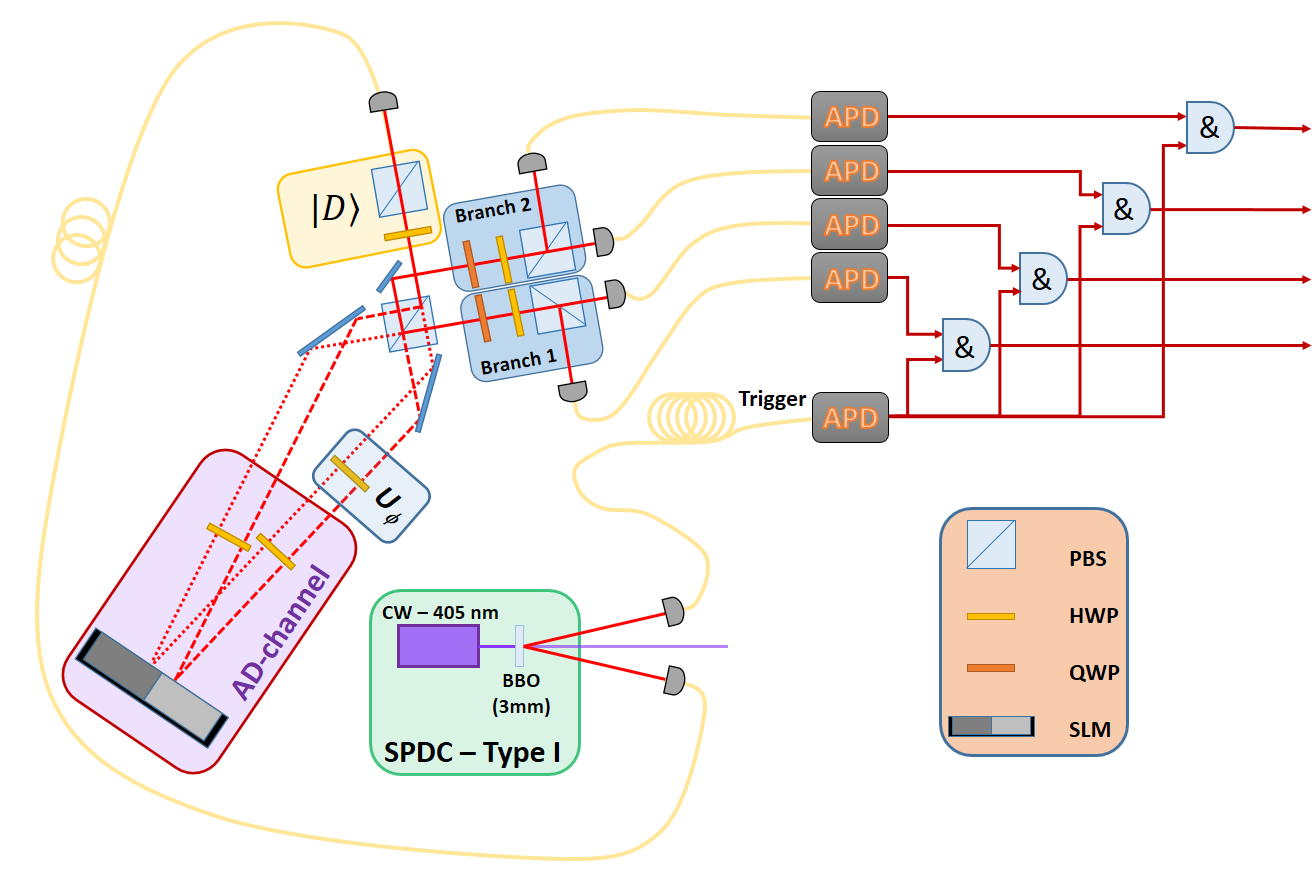}
\caption{Experimental setup. A 405 nm CW diode laser undergoes SPDC on a 3mm BBO Type I crystal. One of the two photons produced is used as a trigger, while the other feeds a Sagnac interferometer in which one of the mirrors is replaced by a SLM. The polarisation of the input state is set on $\vert D\rangle$. The SLM and two of the HWPs at 22.5$^{\circ}$, constituting the AD channel, are used to implement the polarisation-path coupling. A tunable phase mask is applied on the SLM allowing to vary the amount of noise in the channel. A third HWP provides a unitary operation $U_{\phi}$, used to set the phase to be estimated. The polarisation of the output states of the Sagnac are analysed by means a quarter wave plate (QWP), HWP and a PBS. Coincidences between the outputs and the trigger are detected by means of avalanche photodiodes (APD).}
\label{setup}
\end{figure}

For experimental simplicity the ancillary degree of freedom was encoded in the different paths in a Sagnac interferometer, where a Spatial Light Modulator (SLM) replaces one of the mirrors, as schematically depicted in Fig. \ref{setup}. Such a scheme allows for the implementation of an AD channel~\cite{mancino2017}: the H and V polarisation components of the input state are coupled to different paths using a Polarising Beam Splitter (PBS), hence it is possible to simulate a controlled noisy channel by a path-dependent change of the polarisation. Two Half Wave Plates (HWP) and the SLM, which imparts tunable birefringence, implement the polarisation-path coupling. The greyscale image on the SLM is chosen in such a way that each path hits a region corresponding to a different tone. The birefringence is chosen in such a way that the overall transformation leaves the H polarisation unaltered on the clockwise loop; on the counterclockwise loop, a fraction $\eta$ of the initial V component is rotated to H: $\vert V\rangle \rightarrow \left(\sqrt{1-\eta}\right)\vert V\rangle + \sqrt{\eta} \vert H\rangle$. The phase $\phi$ to be estimated has been obtained via a unitary transformation $U_{\phi}=\vert H\rangle\langle H\vert+e^{i\phi}\vert V\rangle\langle V\vert$ by means an additional HWP inside the Sagnac interferometer. Once these two loops are back together on the PBS a new polarisation-path coupling occur: the components that have been left unaffected by the channel are  coherently recombined with the phase $\phi$ on the output Branch 1. The H contribution that originated from the damping emerges separately on Branch 2 (see Fig. \ref{setup}).

The optimal measurement strategy around $\phi\sim\pi$ consists in projecting the polarisation of the photons on Branch 1 on the right (R) and left (L) circular basis, while the other branch can be analysed in any basis since it carries no information on $\phi$; it is a convenient choice to use the H/V basis. 
This choice seconds the symmetry of the output state and gives the following outcome probabilities: $p_R^{(1)}=\frac{1}{4}\left(2-\eta + 2v\sqrt{1-\eta}\sin{\phi}\right)$ and $p_L^{(1)}=\frac{1}{4}\left(2-\eta + 2v\sqrt{1-\eta}\sin{\phi}\right)$, where $v$ is the visibility of our interferometer, for the Branch 1, and $p_H^{(2)}=\frac{\eta}{2}$ and $p_V^{(2)}=0$ for Branch 2. It is demonstrated that this measurement achieves the ultimate limit represented by the quantum Cram\'{e}r-Rao bound (QCRB) for the employed input state; the expression for the corresponding QFI is $F_a=2v^2({1-\eta})/({2-\eta})$, which is always above the single-probe QFI $F_s$~\cite{zixin2016}. 

For phase estimation, we fixed a value for $\eta$ and collected 50 repetitions of four experimental outcome frequencies corresponding to the four channels in Fig.\ref{setup}, in coincidence with the trigger photon. For each repetition, data was accumulated for measurement times of 0.1s, corresponding to a coincidence count rate of nearly 2000 events per acquisition; 50 values of $\phi$ have then been collected. The variance of this sample,  multiplied by the average number of the events, gives the error $(\Delta\phi)^2$, which is expected to converge to the ultimate limit established by the QCRB in the limit of a large number of repetitions; numerical simulations of the ideal case have verified that, for the whole range of $\eta$, the number of events we collected were sufficient to ensure a behaviour close to the asymptotic.

\begin{figure}[t!]
\includegraphics[width=0.5\textwidth]{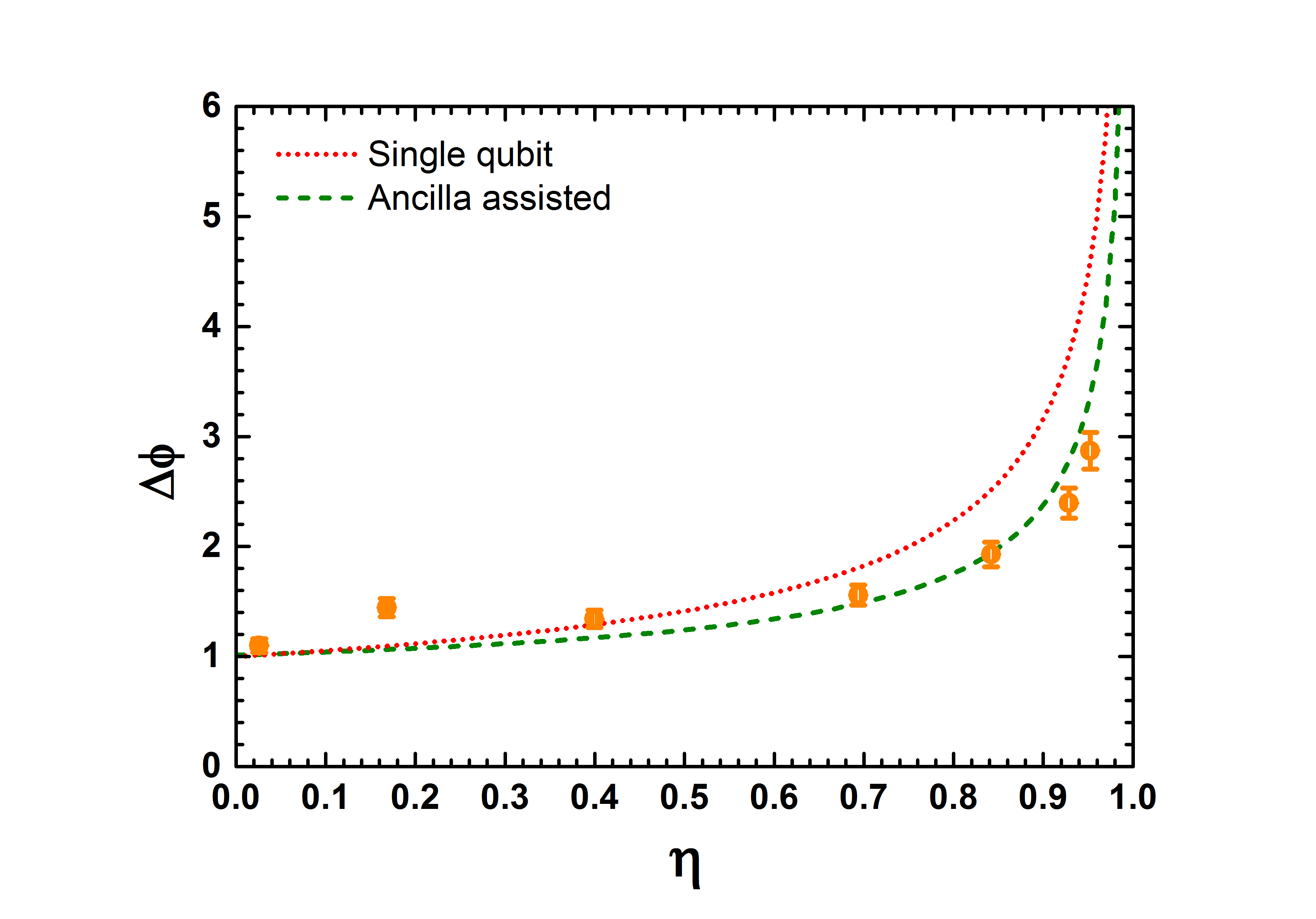}
\caption{Results. We report the behaviour of the standard deviation (SD) of the phase as a function of the channel noise $\eta$. The red dotted curve represents the theoretical SD for the case of a single-qubit estimation measurement. The green dashed line represents the theoretical SD for the ancilla-assisted case. Data, depicted as yellow open circles with error bars, show an effective advantage of the ancilla-assisted strategy in the high-noise regime.}
\label{results}
\end{figure}

Fig. \ref{results} summarises our results: the measured values of $\Delta \phi$ are close to the expected trend of the uncertainty as a function of $\eta$, demonstrating the advantage of adopting the ancilla-assisted strategy in the high-noise regime. At low noise, technical imperfections namely additional noise sources - including non-unit visibility of the interference - prevent to observe any advantage, and make the single-probe strategy more convenient.
 
The transformations of our device can be interpreted as a series of quantum operations on a two-qubit system, one associated to the polarisation, the other to the interferometric path of the single photon. In our experiment the polarisation qubit acts as the probe, while the path qubit represents the ancilla. The action of the first passage on the PBS is then described in these terms by that of a controlled-Not gate, in which the polarisation acts as the control~\cite{fiorentino2005,barbieri2007}. The second passage, instead, allows to implement a measurement projecting on entangled states with a second application of the gate. These considerations, which actually informed the proposal of our experiment~\cite{zixin2016}, clarify the working principle of the protocol at a general level: through the introduction of the ancilla, one engineers the state in such a way that the information on the phase is concentrated in a specific subset of all possible configurations. This subset can be then addressed by a (possibly collective) measurement. The higher signal-to-noise ratio can compensate for the discard of noisy events, and result in an improved QFI.

In our experiment, we can certainly attribute the metrological advantage to the fact that, at the single-particle level, polarisation and path qubits are in a non-separable state. However, the same outcome statistics could be replicated with an attenuated laser, achieving, in principle, a similar improvement of the Fisher information (FI). Indeed, considering polarisation-sensitive loss as one of the possible extensions of amplitude damping to infinite dimensions, introducing additional modes in the interferometer aids the estimation even in situations where entanglement does not play a role. The use of the ancillary modes delivers an improved FI, since polarisation-path coupling remains a coherent operation in this regime as well. The absence of entanglement in this regime does not conflict with the previous observations, since coherent states are known  to be sub-optimal for lossy interferometry~\cite{dorner2009}. It remains as yet an open question whether the optimal states would take advantage of entangled ancillas. But, more importantly, one can take inspiration from this feature of coherent states to look for applications and futher developments in the classical regime as well.

We have demonstrated the advantage of ancilla-assisted protocols in a phase estimation experiment run with single photons, and discussed general considerations on the applicability of this ideas beyond this regime, 
opening to the perspective for looking at this as  an appealing method for phase estimation in noisy environments.

\textit{Note:} During preparation of this manuscript we became aware that similar work was being independently carried out by the group of P. Xue.

We thank M.A. Ricci and F. Somma for discussion, P. Aloe for technical assistance, and F. Sciarrino for the loan of equipment. This work has been funded by the European Commission via the Horizon 2020 programme (grant agreement number 665148 QCUMbER).


\begin{thebibliography}{99}

\bibitem{giovannetti2004} V. Giovannetti, S. Lloyd and L. Maccone, Science 306, 1330-1336 (2004). 

\bibitem{giovannetti2006}  V. Giovannetti, S. Lloyd and L. Maccone, Phys. Rev. Lett. 96, 010401 (2006). 

\bibitem{holland1993} M. J. Holland and K. Burnett, Phys. Rev. Lett. 71, 1355 (1993). 

\bibitem{dorner2009} U. Dorner, R. Demkowicz-Dobrzan\`{s}ki, B. J. Smith, J. S. Lundeen, W. Wasilewski, K. Banaszek and I. A. Walmsley, Phys. 
Rev. Lett. 102, 040403 (2009). 

\bibitem{kacp2010} M. Kacprowicz, R. Demkowicz-Dobrzan\`{s}ki, W. Wasilewski, K. Banaszek and I. A. Walmsley,Nature Photon. 4, 357 (2010). 

\bibitem{knysh2011} S. Knysh, V. N. Smelyanskiy and G. A. Durkin, Phys. Rev. A 83, 021804(R) (2011). 

\bibitem{escher2011} B. M. Escher, R. L. de Matos Filho and L. Davidovich, Nature Phys. 7, 406-411 (2011). 

\bibitem{genoni2011} M. G. Genoni, S. Olivares and M. G. A. Paris, Phys. Rev. Lett. 106, 153603 (2011). 

\bibitem{genoni2012} M. G. Genoni, S. Olivares, D. Brivio, S. Cialdi, D. Cipriani, A. Santamato, S. Vezzoli and M. G. A. Paris, Phys. Rev. A 85, 043817 (2012). 

\bibitem{huttner1996}B. Huttner, A. Muller, J. D. Gautier, H. Zbinden, and N. Gisin Phys. Rev. A 54, 3783 (1996)

\bibitem{pryde2005} G. J. Pryde, J. L. O'Brien, A. G. White, S. D. Bartlett, Phys. Rev. Lett. 94, 220406 (2005)

\bibitem{mosley2006}P. J. Mosley, S. Croke, I. A. Walmsley, S. M. Barnett, Phys. Rev. Lett., 97, 193601 (2006)  

\bibitem{ling2008} A. Ling, A. Lamas-Linares, and C. Kurtsiefer, arXiv:0807.0991, 2008  

\bibitem{baek2008} S.-Y. Baek, S. S. Straupe, A. P. Shurupov, S. P. Kulik, and Y.-H. Kim, Conference on Lasers and Electro-Optics/Quantum Electronics and Laser Science Conference and Photonic Applications Systems Technologies OSA Technical Digest (CD) (Optical Society of America, 2008), paper QFI6; arXiv:0804.3327 (2008).

\bibitem{saunders2012} D. J. Saunders, M. S. Palsson, G. J. Pryde, A. J. Scott, S. M. Barnett, H. M. Wiseman, New J. Phys. 14, 113020 (2012)

\bibitem{altepeter2003} J. B. Altepeter, D. Branning, E. Jeffrey, T. C. Wei, P. G. Kwiat, R. T. Thew, J. L. O'Brien, M. A. Nielsen, and A. G. White, Phys. Rev. Lett. 90, 193601 (2003)

\bibitem{chiuri2011} A. Chiuri, V. Rosati, G. Vallone, S. Padua, H. Imai, S. Giacomini, C. Macchiavello, and P. Mataloni, Phys. Rev. Lett. 107, 253602 (2011)

\bibitem{zhou2011} X.-Q. Zhou, T. C. Ralph, P. Kalasuwan, M. Zhang, A. Peruzzo, B. P. Lanyon, J. L. O'Brien, Nature Communications 2:413 (2011)

\bibitem{layon2009} B. P. Lanyon, M. Barbieri, M. P. Almeida, T. Jennewein, T. C. Ralph, K. J. Resch, G. J. Pryde, J. L. O'Brien, A. Gilchrist, and A. G. White, Nature Physics 5, 134 (2009)

\bibitem{monz2009} T. Monz, K. Kim, W. H\"{a}nsel, M. Riebe, A. S. Villar, P. Schindler, M. Chwalla, M. Hennrich, and R. Blatt, Phys. Rev. Lett. 102, 040501 (2009)

\bibitem{hor2014} M. Hor-Meyll, J. O. de Almeida, G. B. Lemos, P. H. Souto Ribeiro, and S. P. Walborn, Phys. Rev. Lett. 112, 053602 (2014)

\bibitem{demko2014} R. Demkowicz-Dobrza\`{n}ski, and L. Maccone, Phys. Rev. Lett. 113, 250801 (2014).

\bibitem{zixin2016} Z. Huang, C. Macchiavello, and L. Maccone, Phys. Rev. A 94, 012101 (2016).

\bibitem{mancino2017} L. Mancino, M. Sbroscia, I. Gianani, E. Roccia, and M. Barbieri, Phys. Rev. Lett. 118, 130502 (2017). 

\bibitem{fiorentino2005}M. Fiorentino, T. Kim, and F. N. Wong, Phys. Rev. A, 72(1), 012318 (2005).

\bibitem{barbieri2007} M. Barbieri, G. Vallone, P. Mataloni, and F. De Martini, Phys. Rev. A 75, 042317 (2007).


\bibitem{girolami2014} D. Girolami, A. M. Souza, V. Giovannetti, T. Tufarelli, J. G. Filgueiras, R. S. Sarthour, D. O. Soares-Pinto, I. S. Oliveira, and G. Adesso,
Phys. Rev. Lett. 112, 210401 (2014).

\bibitem{roccia2017} E. Roccia, I. Gianani, L. Mancino, M. Sbroscia, F. Somma, M. G. Genoni, and M. Barbieri, arXiv:1704.03327 (2017).

\bibitem{dambrosio2013} V. D'Ambrosio, N. Spagnolo, L. Del Re, S. Slussarenko, Y. Li, L. C. Kwek, L. Marrucci, S. P. Walborn, L. Aolita, and F. Sciarrino, Nat. Comm. 4, 2432 (2013).


\bibitem{altorio2015} M. Altorio, M. G. Genoni, M. D. Vidrighin, F. Somma, and M. Barbieri, Phys. Rev. A 92, 032114 (2015).

\bibitem{altorio2016} M. Altorio, M. G. Genoni, F. Somma, and M. Barbieri, Phys. Rev. Lett. 166, 100802 (2016).

\bibitem{thiel2017}V. Thiel, J. Roslund, P. Jian, C. Fabre and N. Treps, Quantum Sci. Technol. 2, 034008 (2017)

\end{thebibliography}
\end{document}